\documentclass[journal]{IEEEtran}

\usepackage{cite}
\usepackage{amsmath,amssymb,amsfonts}
\usepackage{algorithmic}
\usepackage{graphicx}
\usepackage{textcomp}
\usepackage{longtable}
\usepackage{tabularx,booktabs}
\usepackage{float}
\usepackage{graphicx}
\usepackage{enumerate}
\usepackage{xcolor}
\usepackage{subfigure}
\usepackage{array}
\newcolumntype{L}[1]{>{\raggedright\let\newline\\\arraybackslash\hspace{0pt}}m{#1}}
\usepackage{mhchem}
\usepackage{caption}
\usepackage{multirow} 
\usepackage{latexsym}
\usepackage{url}
\graphicspath{{figures/}}

\ifCLASSINFOpdf
 \else
 \fi
\begin{document}

This work has been submitted to the IEEE for possible publication. Copyright may be transferred without notice, after which this version may no longer be accessible.

\title{State-of-the-Art in Smart Contact Lenses for Human Machine Interaction}

\author{Yuanjie Xia,
        Mohamed Khamis,
        F. Anibal Fernandez,
        Hadi Heidari,
        Haider Butt,
        Zubair Ahmed,
        Tim Wilkinson,
        Rami~Ghannam
\thanks{\textbf{This work has been submitted to the IEEE for possible publication. Copyright may be transferred without notice, after which this version may no longer be accessible.}}
\thanks{Y. Xia, H. Heidari and R. Ghannam are with the School of Engineering, University of Glasgow, UK. e-mail: rami.ghannam@glasgow.ac.uk.}
\thanks {Z. Ahmed is with the 
Institute of Inflammation and Ageing, University of Birmingham, UK. email: z.ahmed.1@bham.ac.uk.}
\thanks {M. Khamis is with the School of Computing, University of Glasgow, UK. email: mohamed.khamis@glasgow.ac.uk.}
\thanks {F. Anibal Fernandez is with the Department of Engineering, University College London, UK. e-mail: a.fernandez@ucl.ac.uk.}
\thanks {T. Wilkinson is with the Department of Engineering, 
University of Cambridge, UK. email: tdw@eng.cam.ac.uk.}
\thanks{H. Butt is with the Department of Mechanical Engineering, Khalifa University,
Abu Dhabi, UAE. e-mail: haider.butt@ku.ac.ae.}
\thanks{Manuscript received XXXX; revised XXXX}}


\maketitle

\begin{abstract}
Contact lenses have traditionally been used for vision correction applications. Recent advances in microelectronics and nanofabrication on flexible substrates have now enabled sensors, circuits and other essential components to be integrated on a small contact lens platform. This has opened up the possibility of using contact lenses for a range of human-machine interaction applications including vision assistance, eye tracking, displays and health care. In this article, we systematically review the range of smart contact lens materials, device architectures and components that facilitate this interaction for different applications. In fact, evidence from our systematic review demonstrates that these lenses can be used to display information, detect eye movements, restore vision and detect certain biomarkers in tear fluid. Consequently, whereas previous state-of the-art reviews in contact lenses focused exclusively on biosensing, our systematic review covers a wider range of smart contact lens applications in HMI. Moreover, we present a new method of classifying the literature on smart contact lenses according to their six constituent building blocks, which are the sensing, energy management, driver electronics, communications, substrate and the I/O interfacing modules. Based on recent developments in each of these categories, we speculate the challenges and opportunities of smart contact lenses for human-machine interaction. Moreover, we propose a novel self-powered smart contact lens concept with integrated energy harvesters, sensors and communication modules to enable autonomous operation. Our review is therefore a critical evaluation of current data and is presented with the aim of guiding researchers to new research directions in smart contact lenses.
\end{abstract}

\begin{IEEEkeywords}
Contact Lenses, Human Machine Interaction, Human Computer Interaction, Wearables.
\end{IEEEkeywords}

\IEEEpeerreviewmaketitle

\section{Introduction}
\IEEEPARstart{C}{ontact} lenses are thin lenses which could be placed directly on the surface of human eyes. They have historically been used for vision correction and they were first developed by Adolf Fick in 1887 to correct astigmatism \cite{lahdesmaki2010possibilities}. These lenses have now undergone a series of innovations and are gaining popularity due to recent developments in materials engineering and microfabrication. Currently, these innovative platforms are mainly being considered for a plethora of medical applications since they are noninvasive \cite{park2018soft} and and can be used to deliver medicines \cite{xinming2008polymeric}. In comparison to traditional implantable devices, contact lenses do not require surgery and are capable of monitoring bio-markers continuously. This is especially true since tear fluid carries philological bio-markers that can be used to monitor the health of its wearer  \cite{kim2020electrochromic}. 

Advancements in smart contact lenses was impeded by the ability to communicate the `sensed' data wirelessly. In fact, this technological challenge hindered the development of smart contact lenses during the past two decades. However, advanced microfabrication technologies enabled miniature electronic components to be integrated on a contact lens. Contact lenses can now integrate sensors, driver electronics, energy modules, antennae and I/O terminals on the same platform. Progress in microelectronics packaging technology as well as nanofabrication and nanocommunications has therefore resulted in multiple potential applications for smart contact lenses. For example, in addition to their use for vision correction, medical staff could potentially monitor diabetes patients in real-time using these smart contact lenses \cite{gonzalez2016transparent,quintero2020artificial,raducanu2020artificial}.  

\begin{figure}[h] 
	\centering  
	\subfigure[]{
		\label{fig.1 a}
		\includegraphics[width=0.9\linewidth]{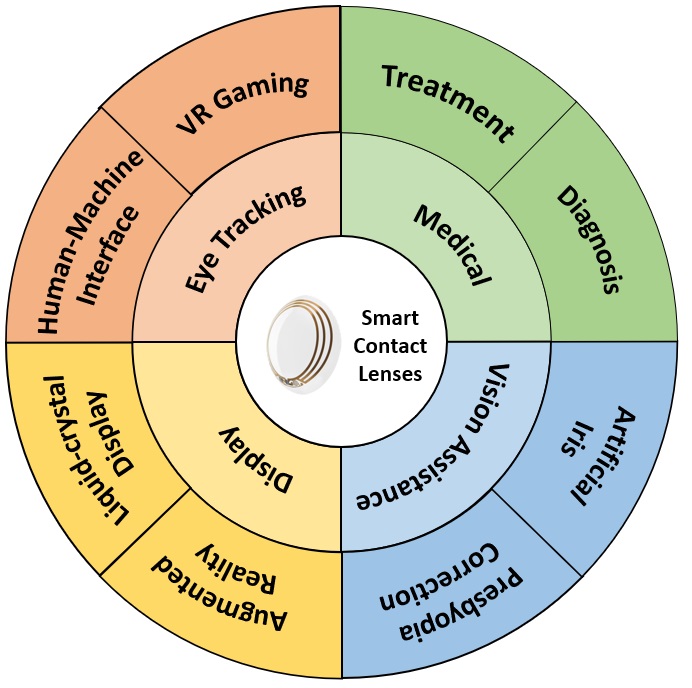}}

	\subfigure[]{
		\label{fig.1 b}
		\includegraphics[width=0.9\linewidth]{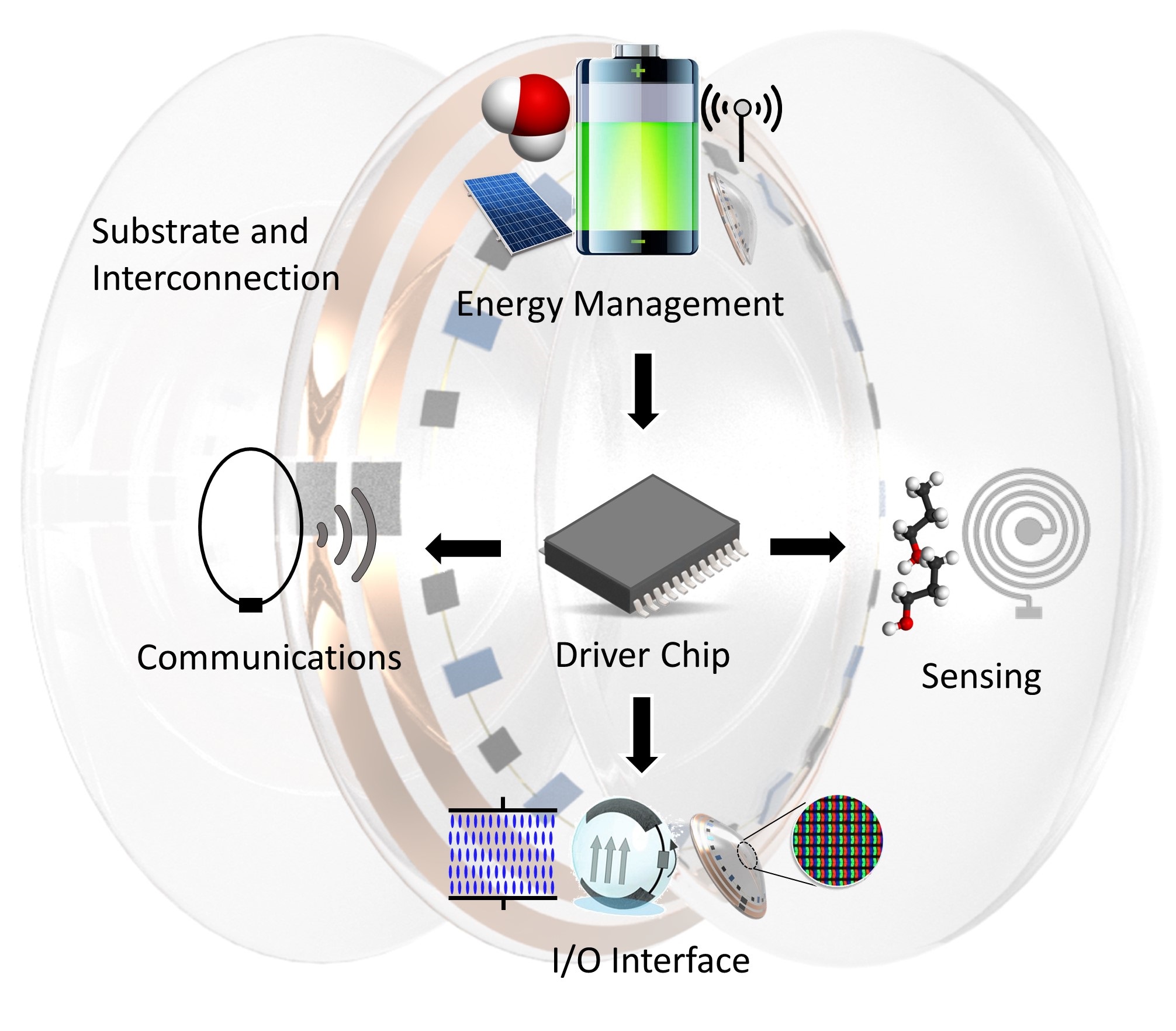}}
	\label{fig.1}
\caption{(a) Range of contact lens applications for Human-Machine Interaction (HMI) that include medical, eye tracking, display and vision assistance. (b) Constituent building blocks of smart contact lenses, which include the Energy Management, Communications, Driver Chip, Sensing, I/O Interface and the Substrate \& Interconnection modules.}
\end{figure}

In addition to the healthcare field, there are a variety of promising smart contact lens applications, as shown in fig. \ref{fig.1 a}. According to our investigations,
applications of existing smart contact lens could be divided into four main categories, which are: (a) eye tracking, (b) healthcare (medical), (c) displays and (d) vision correction. We will discuss recent progress in each of these application areas in section VI of our article. Moreover, we have identified five constituent building blocks of smart contact lenses, which are the (a) Sensor, (b) Energy, (c) Driver Chip, (d) Communications, (e) Input/Output (I/O) Interface and (f) Substrate \& Interconnection modules. We chose to discuss recent developments in smart contact lenses according to these five building blocks. As is shown in fig. \ref{fig.1 b}, the Energy module is concerned with power harvesting and storage, while the Driver Chip module provides and regulates this energy, ensuring all electronic modules operate safely. Furthermore, we included a fifth module called ``Substrate \& Interconnection'', which hosts and connects all these electronic modules together. 

First, in section II we demonstrate how our eyes can be used for HMI applications from a physiological and technical perspective. Next, in section III of the paper we discuss the State-of-the-Art in review papers on smart contact lenses. In section IV we describe our methodology in compiling relevant research articles that match our search criteria. Subsequently, we discuss and analyse results from our search in section V. Moreover, based on these recent developments and to ensure that smart contact lenses can be used for a variety of HMI applications, we propose an autonomously driven concept contact lens in section VI. Finally, the paper ends with concluding remarks and a proposed outlook in section VII.

\section{Using our Eyes for HMI}
HMI is concerned with how people and automated systems interact and communicate with each other. This has long ceased to be confined to just traditional machines in industry and now also relates to computers, digital systems or devices for the 'internet of things (IoT) \cite{chen2018securing}. More and more devices are connected and automatically carry out tasks in the background. Operating all of these machines, systems and devices need to be intuitive and must not place excessive demands on users. Therefore, smooth communication between people and machines requires user-friendly interfaces. A contact lens is one of those interfaces that could broaden the potential applications and facilitate HMI. The purpose of this section is to demonstrate that eye movements could be used to facilitate HMI.

\subsection{Physiology of the Eye}
\begin{figure}[htb]
\centering
\includegraphics[scale=1.3]{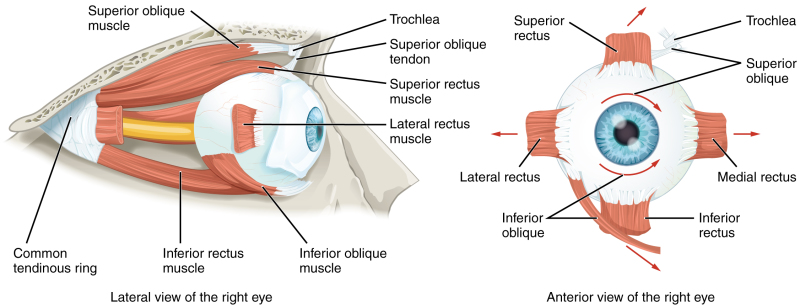}
\caption{Schematic diagram of the eye, which shows six essential muscles that are responsible for eye movements (adapted from \cite{eyemodel} © 2021, StatPearls Publishing LLC.)}
\label{fig:eye_diagram}
\end{figure}

Before reviewing the range of contact lens technologies that can be used for HMI, it is important to understand the basic physiology of the eye, how it moves and what are the muscles responsible for this movement. The eye is a spherical organ with a mean diameter of around 24 mm. As demonstrated in figure \ref{fig:eye_diagram}, eye rotation is controlled by six ocular muscles, which are the superior rectus, inferior rectus, lateral rectus, medial rectus, the superior oblique and the inferior oblique. The contraction of these six eye muscles result in eye movements \cite{wright2003anatomy}. There are mainly three reasons for eye movements, which are reducing image motion, positioning the fovea and avoiding double vision \cite{carpenter1988movements,leigh2015neurology}. Moreover, there are six types of basic eye movements, which include abduction, adduction, supraduction, infraduction, incyclotorsion and excyclotorsion \cite{kohli2018cranial}. In abduction and adduction, the cornea moves away from and towards the midline, respectively. Similarly, supraduction and infraduction means that the cornea moves upwards and downwards, respectively. In addition to the previous four movements, the eye can rotate clockwise or counterclockwise thanks to the superior oblique and inferior oblique muscles, respectively. In fact, every complex eye movement is made up of these six basic eye movements, which could be controlled by humans consciously.

\subsection{Advantages of Eyes for HMI}
HMI applications require an input device, which may involve a touch-screen device, touch pad or keyboard \cite{soukupova2016eye}. However, most of these input devices require manual manipulation. Moreover, since our eyes are used to collect the necessary visual information from these machines, we can also use them to control and interact with these machines for HMI application. Consequently, hardware and software can be developed to detect eye movements and gestures such as gaze and blinking to facilitate HMI. In fact, there are a variety of methods in the literature that can be used for gaze and eye tracking, which include computer vision \cite{cheung2015eye,corcoran2012real}, magnetic sensing \cite{tanwear2020spintronic} and laser sensing \cite{massin2020smart}. 

Therefore, in comparison to traditional input devices, eye-related gesture recognition and tracking avoids physical contact and benefits people with impaired movement. For example, previous studies in the literature confirmed that eye blinking could be used to facilitate HMI \cite{krolak2012eye,morris2002blink}. Here, `blinking' is a semi-automatic reflex function that is controlled by two muscles, which are the levator palpebrae superioris and the orbicularis oculi \cite{porter1989morphological}. People sometimes blink consciously as a form of body language or gesture, which is interpreted as `winking'. Therefore, by combining blink detection, gaze tracking and eye movement, our eyes could be used as an effective and user-friendly interface for HMI applications. 

\begin{figure*}[htb]
\centering
\includegraphics[scale=0.6]{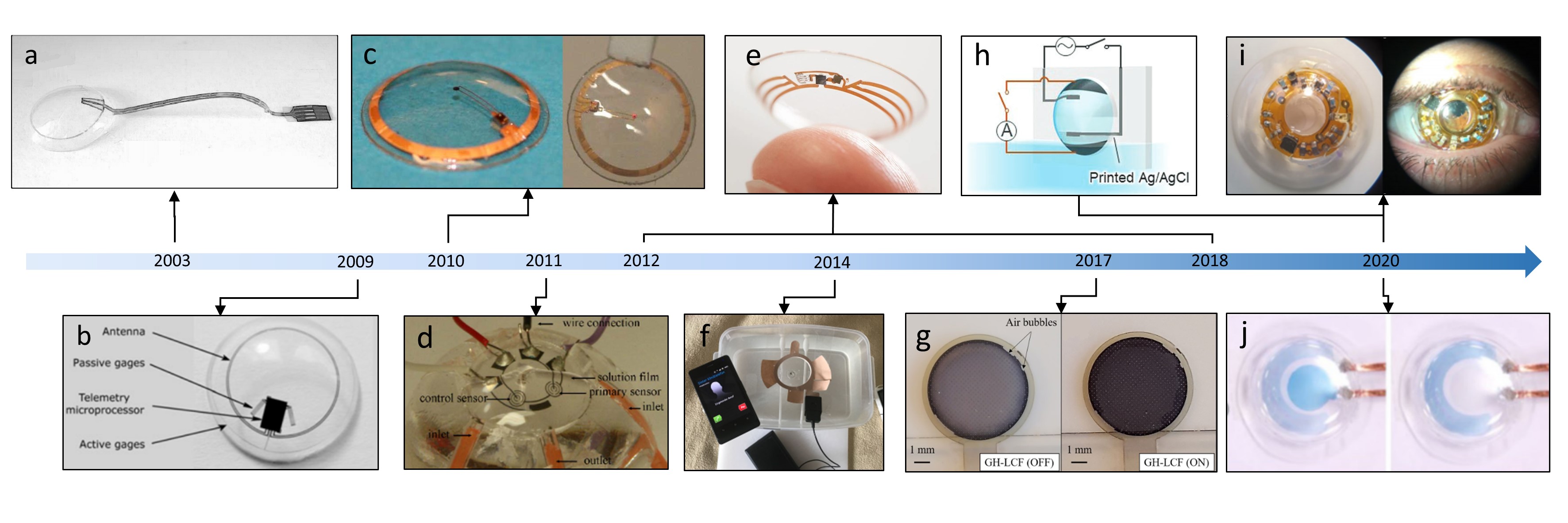}
\caption{Timeline of smart contact lens development. (a) Earliest smart contact lens developed by M. Leonardi \textit{et al.} in 2003 (adapted from \cite{leonardi2003soft}). (b) Subsequent advancements in the micro-fabrication industry led to the development of more sophisticated contact lenses 6 years later. For example, the first contact lens designed for intraocular pressure (IOP) monitoring (reproduced with permission from \cite{leonardi2009wireless} © 2008 Acta Ophthalmol). In (c), an antenna was integrated on a contact lens for energy harvesting. The harvested power was used to power a single-pixel LED (Reproduced with permission from \cite{pandey2010fully} © 2010 IEEE). (d) The first contact lens designed for glucose monitoring (reproduced with permission from \cite{yao2011dual}. ©2011 IEEE). (e) Google aimed to commercialize contact lenses for glucose measurement, but the project was discontinued due to inaccuracies when using tear glucose to predict blood glucose (reproduced with permission from \cite{choi2017smart} © 2017 American Chemical Society). (f) SENSIMED also developed and commercialized contact lenses for IOP sensing (reproduced with permission from \cite{rabensteiner2018influence} © 2018 BMC Ophthalmology). (g) The first contact lens artificial iris design, which used liquid crystals (LCs) for changing the transparency of the contact lens (reproduced with permission from \cite{quintero2020artificial} © 2020 Nature). The next images show most recent smart contact lens applications for (h) dry eye treatment (reproduced with permission from \cite{kusama2020self} © 2019 WILEY-VCH), (i) gaze tracking (reproduced with permission from \cite{khaldi2020laser} © 2020 Nature) and (j) hazard perception (reproduced with permission from \cite{kim2020electrochromic} © 2020 Elsevier).}
\end{figure*}

\section{State of the Art}
Over the past two decades, major advances in electronics have allowed researchers to design and integrate miniature electronic systems into contact lens platforms. Smart contact lenses became an increasingly popular topic in electronic wearable devices. In previous studies, researchers have proposed and realized contact lens for health monitoring, gaze tracking and many other applications. The purpose of this section is to survey the latest review articles of contact lens technology. Additionally, we will also conclude the limitations of previous review paper and show how our work could address the limitations.

Farandos \textit{et al.} published the first review article on smart contact lenses in 2015 \cite{farandos2015contact}. They confirmed that contact lenses are attractive platforms for health monitoring applications in comparison to other wearable technologies due to their `minimalistic nature'. Their manuscript started with background information regarding the physical characteristics of the eye, followed by an analysis into tear fluid composition.  They reviewed the state-of-the-art in contact lens sensor fabrication, detection techniques as well as the energy harvesting and readout systems. They also analyzed the market for contact lenses and FDA regulatory requirements for commercialization of contact lens sensors. Their manuscript focused on reviewing advances in monitoring intraocular pressure (IOP) and tear glucose, which were the two main application areas of smart contact lenses at that time. Their paper mentioned that commercial contact lens products require sufficient selectivity, sensitivity, reproducibility and could ensure patient compliance. They also mentioned that treatment capabilities could be integrated in contact lens platform. For future development, they highlighted three main challenges with contact lenses, which were energy harvesting, wireless connectivity and sensor stability.

Moreover, in 2018 Tseng \textit{et al.} evaluated different kinds contact lenses for biosensing applications \cite{tseng2018contact}. Their article focused on reviewing the different methods for using tear fluid as a detection medium for different diseases such as cancer, ocular disorders and diabetes. They reviewed previous manuscripts about tear content, and evaluated both active as well as passive sensors for tear content sensing. Their review paper predicted that future smart contact lenses would lead to better personalized medical treatment. Similar to the review article by Farandos \textit{et al.}, they again highlighted that contact lens stability and sensor data repeatability is a challenge. Other challenges that were mentioned were wearer comfort and cost.

In 2020, Kim \textit{et al.} published another review article on smart contact lenses for biosensing applications that focus on diseases diagnosis \cite{kim2020recent}. Their review analysed both physical and chemical sensors. They also briefly introduced state-of-the-art in contact lenses for drug delivery, data transmission and power storage. They mentioned that further research is necessary to fully appreciate what biomarkers can be detected in tear fluid. They also indicated that an optimisation of sensor accuracy is necessary, since this depends on the nature of the disease to be detected and some diseases require higher sensing accuracy than others. Moreover, they raised concerns regarding sensor stability and bio-compatibility. However, they concluded that smart contact lenses could enable noninvasive health monitoring and diagnosis based on people’s tear fluids, which is likely to be used more extensively. 

Based on the above, state-of-the-art review articles on smart contact lenses have focused on biosensing applications. In this manuscript, we aim to systematically review the range of contact lens technologies that facilitate HMI. These technologies include biosensing and other relevant technologies, such as liquid crystal displays and augmented reality. In this context, our work goes beyond existing review articles in this field, since it aims to showcase how contact lenses have supported HMI and have been used in applications other than the medical field. 

\section{Methodology}
In this section, we define our research methodology in collecting and synthesizing evidence on smart contact lenses using clearly defined criteria. Academic journal articles and conference papers were chosen for review, considering their relatively high impact. Web of Science, which was one of the largest academic database, was used for identifying the articles using certain keywords to confine the results. In addition, the keywords that were used to the searching query are shown in table \ref{table:Descriptors}. The (AND) Boolean operator was used to connect these descriptors. All manuscripts that contain the aforementioned keywords were included in our review. 

Similar to the methodology described in \cite{ghannam2020implantable}, we first defined the research questions and the inclusion criteria of our search. Second, we selected relevant manuscripts that met these criteria. Third, we analyzed and interpreted our search results. In this case, we defined the following research questions (RQ):

\begin{enumerate}
  \item RQ1: What are the different ways in which contact lenses have been used to facilitate HMI?
  \item RQ2: What are the device architectures?
  \item RQ3: Where is the research activity on smart contact lenses mainly located? 
  \item RQ3: What are the challenges and what is the future outlook?
\end{enumerate}

Based on the above questions, the following inclusion criteria (InC) were defined:

\begin{enumerate}
  \item InC 1: Articles written in English;
  \item InC 2: Articles matching the definitions and descriptors mentioned in table \ref{table:Descriptors}.
\end{enumerate}

\begin{table*}[h!]
	\small\centering
	\caption{DESCRIPTORS AND SYNONYMS USED FOR OUR SEARCH.}
    \begin{tabularx}{\textwidth}{ p{2cm}|p{11cm}|p{4cm} }
       	\specialrule{2.5pt}{3pt}{3pt}
		Descriptor & Definition & Synonyms  \\ 
		\specialrule{2.5pt}{3pt}{3pt}
Contact Lens & 
Contact lenses are thin lenses placed directly on the surface of the eyes \cite{moreddu2019contact}& 
Contacts, soft lenses and extended-wear lenses  \\
Smart & 
A smart device is able to do many of the things that a computer does \cite{smart} & 
Autonomous  \\
Electronic & 
An electronic device has transistors or silicon chips which control and change the electric current passing through the device \cite{electronics} & 
Electronics \\
Wireless & 
Wireless technology uses radio waves rather than electricity and therefore does not require any wires \cite{wireless} & 
RF powered\\
		\specialrule{2.5pt}{3pt}{3pt}
	\end{tabularx}
	\label{table:Descriptors}
\end{table*}

Having defined our research questions as well as our inclusion criteria, we defined our approach in searching for the relevant literature. We used `Web of Science' portal for our search with the descriptors: smart contact lens, electronic contact lens, wireless contact lens and sensing contact lens. A forward and backward search was also performed via 'Google scholar' to make sure that all literature has been captured.

\section{Results and Discussion}
In this section, we analyze the literature on smart contact lenses for HMI applications. In total, 83 manuscripts satisfied our search criteria, which were described in the previous section. The majority of these manuscripts (approximately 61\%) were published during the past 5 years. 

\begin{figure}
\centering
\includegraphics[width=\linewidth]{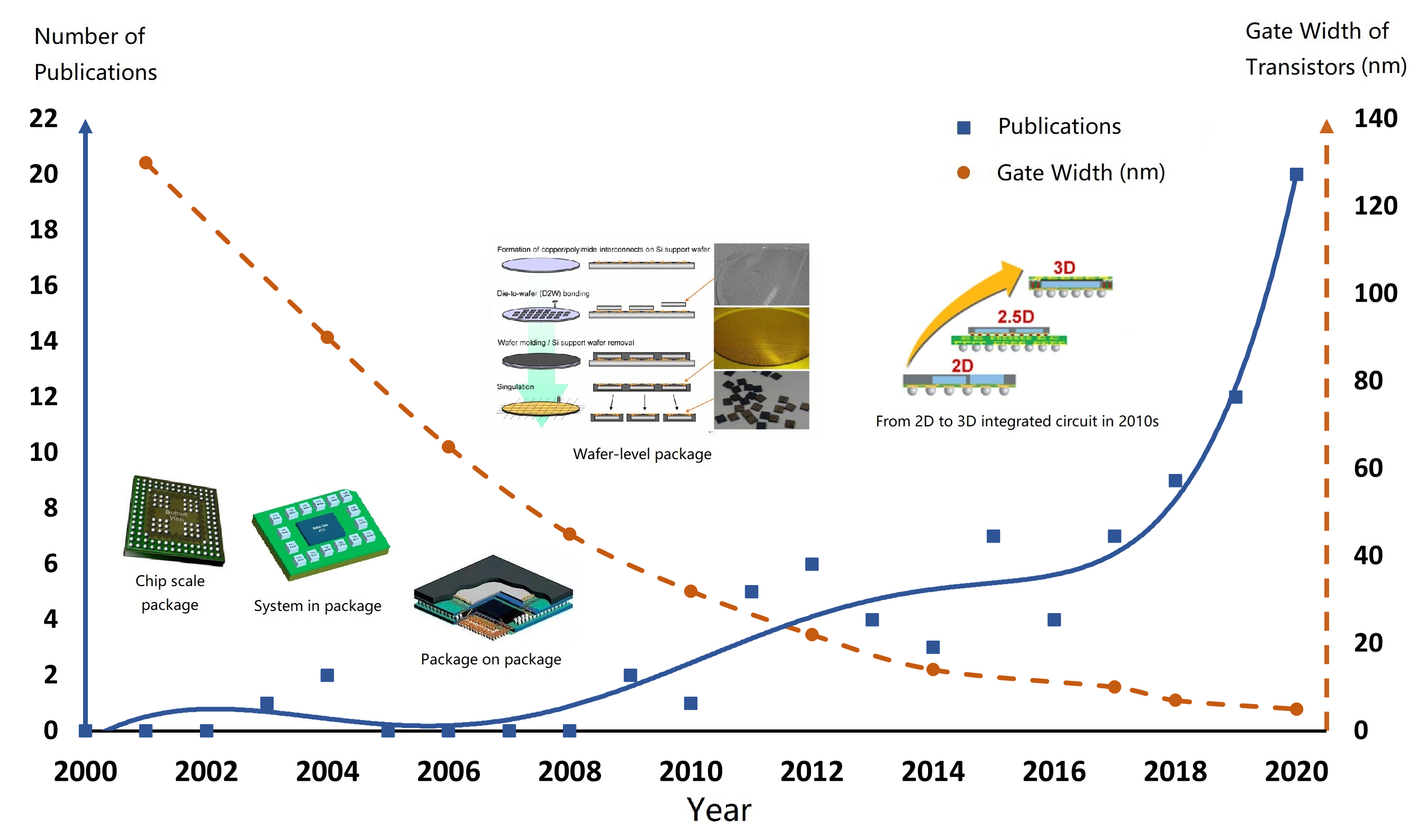}
\caption{Comparison between the publications activity on smart or electronic contact lenses with CMOS gate width dimensions. A rise in publications activity started in 2008, which corresponded to the 45 nm gate width.}
\label{PublicationsActivity}
\end{figure}

Accordingly, we noticed an onset of research interest on smart or electronic contact lenses from 2008, as shown from figure \ref{PublicationsActivity}. This is just one year after the emergence of the first Apple iPhone (2007). It is also just four years after the beginning of the nanometer sized transistors. The year 2008 also signifies an important milestone in the transistor industry, since it marks the emergence of the 45 nm node technology, thus paving the way for advanced processors such as the atom and the core i7 \cite{kuhn2008managing}. Such critical technological advancements had an important role in the development of electronic components on contact lens system. We will discuss the range of technologies in each of these sections. In fact, 50 articles have been published since 2015, which accounts for 73.5\% of all the  publications on smart or electronic contact lenses. We also observe an obvious rise in publications in this field from 2000 to 2020.

\begin{figure}[htb]
\centering
\includegraphics[width=3.4in]{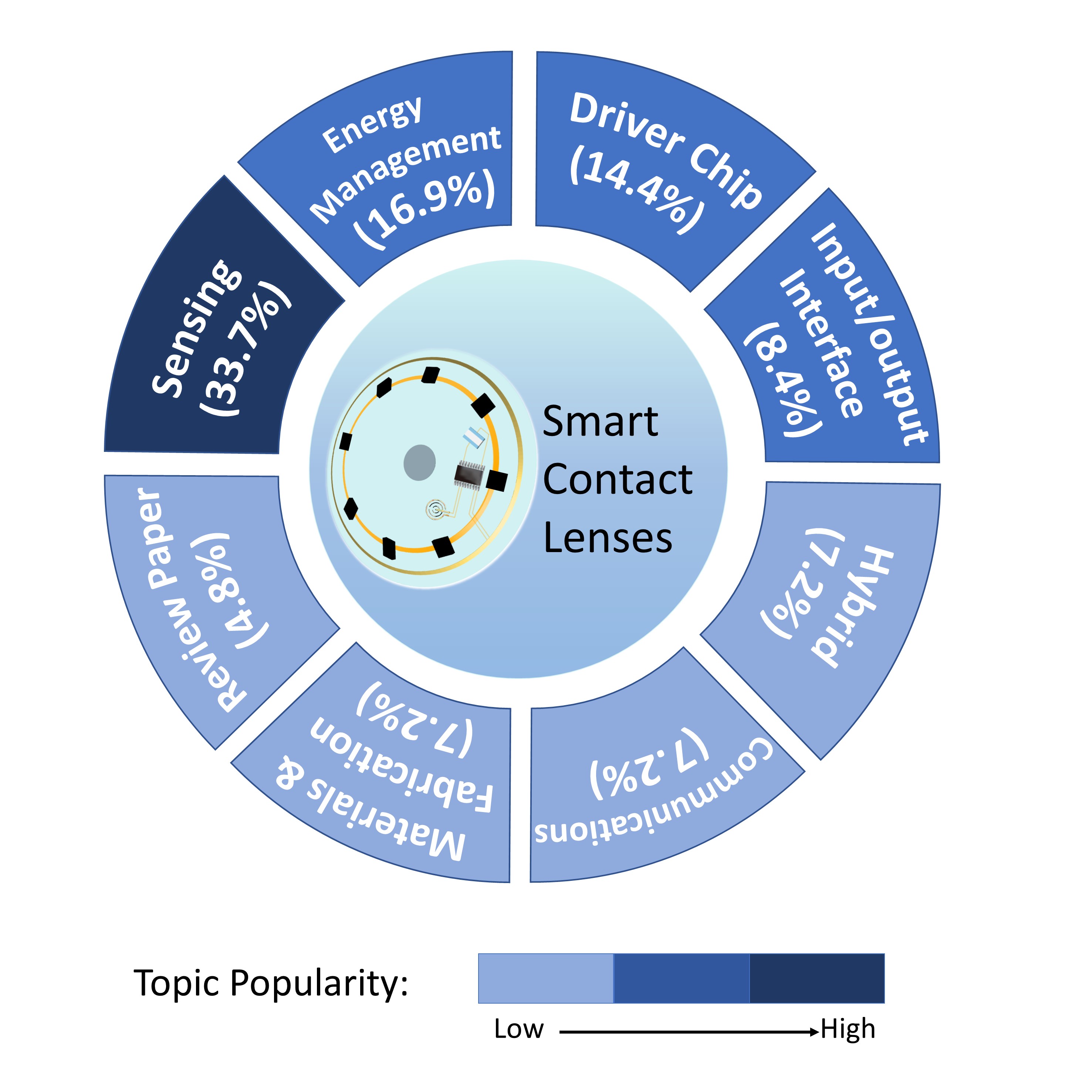}
\caption{Different components and building blocks of a smart contact lens. These are the Sensing, I/O Interfacing, Communications, Driver Chip and Substrate \& Interconnecting modules. A hybrid category was introduced to include smart contact lens developments in more than one building block. Our search also included four review papers. According to our search, the majority of the literature has focused on developments in the sensing, energy and driver chip modules, with fewer publications on the Communications and well as the I/O Interfacing modules.}
\label{fig:Categories}
\end{figure}

Based on gathered data, recent developments in smart contact lenses were categorised according to their six previously mentioned building blocks. About 7.2\% of those manuscripts reported developments in more than one building block and a "Hybrid" category was 
introduced, as shown in figure \ref{fig:Categories}. Furthermore, almost a third (33.7\%) of all the literature was devoted to the ``Sensing'' category, whereas roughly another third reported developments in the ``Energy'' and ``Driver Chip'' modules (16.9\% and 14.4\% of manuscripts, respectively). With the exception of the ``Hybrid'' papers and the four Review papers, a fifth of all the literature reported developments in the remaining four building blocks of smart contact lenses. Latest developments in each of these categories will be discussed in this section. Moreover, details regarding the application areas of smart contact lenses will be discussed in Section VI. 

\begin{figure*}[h!]
\centering
\includegraphics[scale=0.65]{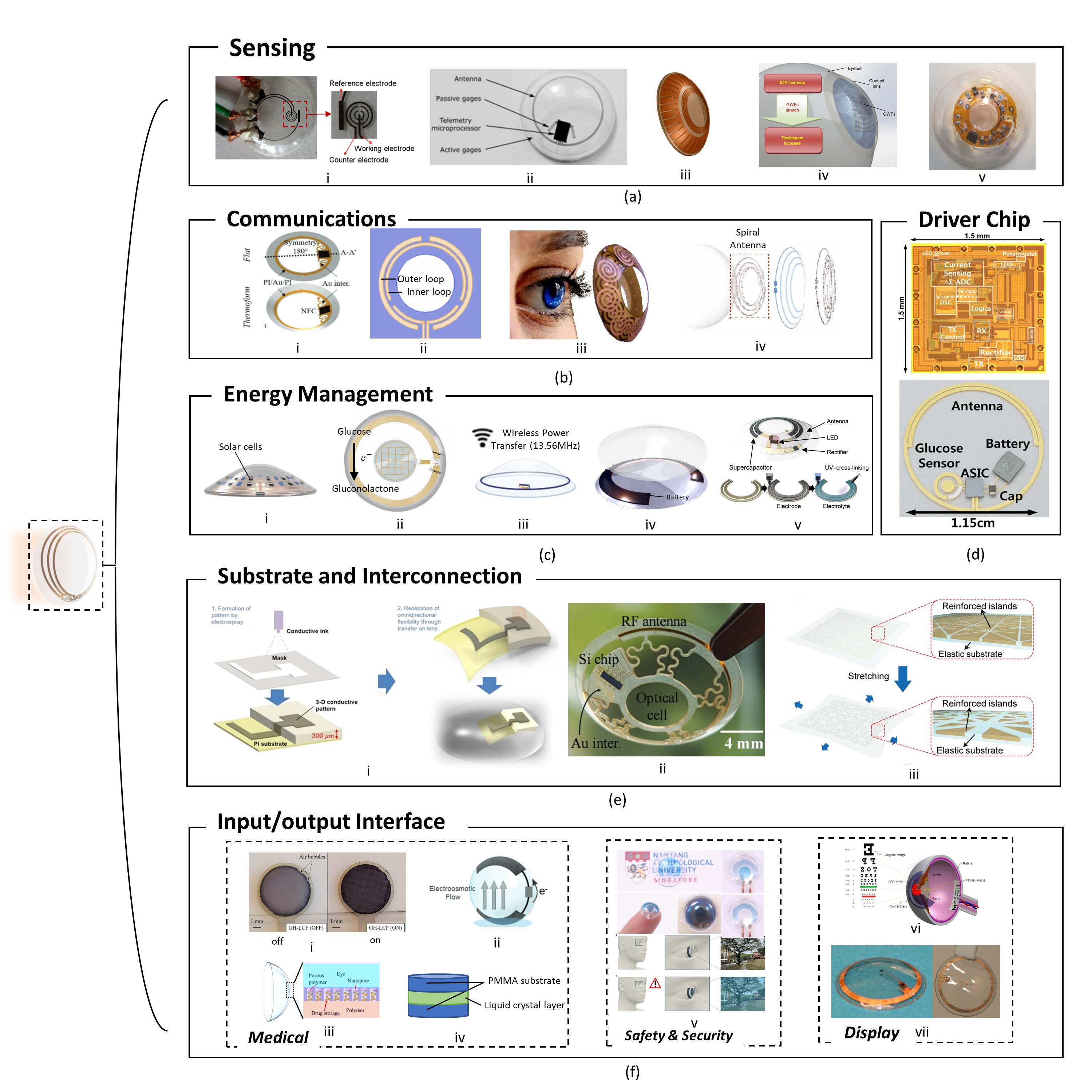}
\caption{Essential building blocks of a smart contact lens. We divided these into the following building blocks: (a) Sensors: (i) Glucose sensing (reproduced with permission from \cite{yao2011contact} ©2011 IEEE), (ii) IOP sensing by Leonardi \textit{et al.} (reproduced with permission from \cite{leonardi2009wireless} © 2008 Acta Ophthalmol), (iii) IOP sensing by Chen \textit{et al.} (adapted from \cite{chen2013capacitive}), (iv) IOP sensing by Zhang \textit{et al.} (reproduced with permission from \cite{zhang2019high} © 2019 Nature) and (v) Eye tracking (reproduced with permission from \cite{khaldi2020laser} © 2020 Nature). Subfigure (b) demonstrates different types of antenna design including: the loop antennae as demonstrated by (i) Vasquez \textit{et al.} (reproduced with permission from \cite{vasquez2020near}, (ii) Chiou \textit{et al.} (reproduced with permission from \cite{chiou2016toward}; the spiral antenna as demonstrated by (iii) Chen \textit{et al.} (reproduced with permission from \cite{chen2017warpage}, and (iv) Kim \textit{et al.} (reproduced with permission from \cite{kim2017wearable}. Subfigure (c) demonstrates different types of energy modules including: energy harvesters, as demonstrated by (i)  Lingley \textit{et al.} \cite{lingley2012contact}, (ii) Falk \textit{et al.} \cite{falk2012biofuel} and (iii) Takamatsu \textit{et al.} \cite{takamatsu2019highly} © 2019 WILEY-VCH). Moreover, power storage methods in contact lenses were demonstrated by (iv) Lee \textit{et al.} \cite{lee2018scalable} and (v) Parking \textit{et al.} \cite{park2019printing} © 2019 American Association for the Advancement of Science. Subfigure (d) shows a driver chip design by Jeon \textit{et al.} \cite{jeon2020smart} © 2019 IEEE. Subfigure (e) demonstrates innovative substrate and interconnection design for smart contact lenses (\cite{kim2018three,vasquez2017stretchable,park2018soft} © 2018 American Chemical Society, © 2017 WILEY-VCH, © 2018 American Association for the Advancement of Science). Subfigure (f) demonstrates input/output terminal block for applications that include, (i) an artificial iris \cite{quintero2020artificial} © 2020 Nature, (ii) dry eye treatment (reproduced with permission from \cite{kusama2020self} © 2019 WILEY-VCH), (iii) drug delivery (adapted from \cite{song2019multifunctional}), (iv) presbyopia treatment (adapted from \cite{milton2014electronic})), (v) safety \& security (reproduced with permission from \cite{kim2020electrochromic} © 2020 Elsevier B.V.), (vi) display demonstrated by Pandey \textit{et al.} (reproduced with permission from \cite{pandey2010fully} © 2020 SPIE (v) display demonstrated by Chen \textit{et al.} \cite{chen2020foveated} © 2010 IEEE).}
\label{fig:SCL_ModuleExamplesfromLiterature} 
\end{figure*}

\subsection{Sensing}
Most contact lenses involve a sensing module, especially since they are widely regarded as a non-invasive platform that could be used to measure vital human signs, such as glucose levels and intraocular pressure (IOP). In recent studies, researchers developed contact lenses for gaze tracking and electroretinogram (ERG) measurement. Common sensor types and applications are listed in Table \ref{table:sensors}. In this section, we will discuss latest sensor developments in contact lens design.

\begin{table*}[h!]
	\small\centering
	\caption{Range of sensor technologies for smart or electronic contact lenses.}
    \begin{tabularx}{\textwidth}{ p{5cm}|p{3cm}|p{6cm}|p{2cm} }
       	\specialrule{2.5pt}{3pt}{3pt}
		Sensor Type & Application & Reference & Percentage \\
		\specialrule{2.5pt}{3pt}{3pt}
		Capacitance Sensor &  \multirow{3}{*}{IOP measurement} & \cite{huang2013contact,song2019multifunctional,rabensteiner2018influence,maeng2020photonic,kim2017wearable,hsu2015rfid,cheng2013rectenna,yeh2015toward,chiou2015addressable,chen2013capacitive,chiou2017towards,chiou2015capacitor,luo2013dual} & \multirow{3}{*}{} \\ \cline{1-1} \cline{3-3}
		Resistance sensor  &  &  \cite{leonardi2004first,zhang2019high,leonardi2003soft,leonardi2009wireless}   &  52.9\%    \\ \cline{1-1} \cline{3-3}
		Inductor sensor   &    &   \cite{tseng2012design}    &     \\ \hline
		Electrochemistry sensor &  Glucose measurement & \cite{hayashi2019design,lahdesmaki2010possibilities,kim2020wireless,jeon2020smart,liao20113,jeon2019143nw,gonzalez2016transparent,yao2011contact,yao2011dual,yao2012soft,park2018soft}  & 32.3\%  \\ \hline
		Laser sensor  & Eye track & \cite{massin2020development,massin2020smart,khaldi2020laser,tanwear2020spintronic,kim2004wireless}          & 14.7\%       \\ \hline
		Graphene electrode array  & ERG measurement & \cite{yin2018soft}          & 2.9\%       \\
		\specialrule{2.5pt}{3pt}{3pt}
	\end{tabularx}
	\label{table:sensors}
\end{table*}

\subsubsection{Glucose Monitoring}
Yao \textit{et al.} demonstrated a wireless health monitoring contact lens for detecting glucose levels in tear \cite{yao2011contact}. Their design could work as a noninvasive and continuous glucose monitor for diabetes, as shown in fig. \ref{fig:SCL_ModuleExamplesfromLiterature}a i. In fact, their contact lens uses enzyme-based electrochemical glucose sensors due to its high selectivity and efficiency. With the glucose oxidase catalysing the reaction between glucose and oxygen in free space to form hydrogen peroxide and gluconolactone \cite{heller2008electrochemical}. Subsequently, the hydrogen peroxide will be detected by electrodes after decomposition.

\subsubsection{IOP Monitoring}
In addition to glucose monitoring, researchers have demonstrated contact lenses that are capable of monitoring IOP directly and continuously. High IOP could be regarded as a driving element for glaucoma, which is a degenerative condition of the posterior pole of the eye and can lead to irreversible blindness. Hence, IOP monitoring contact lenses could prevent the development and progression of this disease \cite{davanger1991probability}. Therefore, an IOP sensor design was devised by Leonardi \textit{et. al} \cite{leonardi2009wireless}. Their IOP sensor relies on mechanical changes caused by raised IOP levels, which lead to a variation in capacitance, as demonstrated in fig. \ref{fig:SCL_ModuleExamplesfromLiterature}a ii.

Furthermore, Zhang \textit{et. al} designed an IOP sensor based on graphene woven fabrics \cite{zhang2019high}. They observed that a variation of IOP causes a change in readout voltage when the sensor current is constant, which implies a change in sensor resistance. Their design is demonstrated in fig. \ref{fig:SCL_ModuleExamplesfromLiterature}a iv. However, their contact lens was characterised for a range of voltages from 0-10 V, with a particular focus on the linear region between 6-10 V, which may be too high for a practical wearable application. In addition, they also claim that this sensing contact lens is low-cost and disposable, which makes it a promising platform for point-of-care medical treatment.

\subsubsection{Gaze Tracking} 
Apart from IOP and glucose sensing, researchers have also investigated the feasibility of developing sensors for gaze tracking. In comparison with the video-based method, eye tracking smart contact lenses are more convenient for constrained environments. For example, Massin \textit{et al.} purposed an eye tracking contact lens using encapsulated photodetectors \cite{massin2020development,massin2020smart}. An infra-Red (IR) source emitter was mounted on an external glasses frame. Photodetectors were illuminated by the external source and eye movement was detected via a variation in the amount of light that the photodetectors register. A similar design was also purposed by Khaldi \textit{et al.}, where an IR emitter was mounted on a smart contact lens that detected eye movements using an external IR camera \cite{khaldi2020laser}, as demonstrated in fig. \ref{fig:SCL_ModuleExamplesfromLiterature}a v. 

Moreover, instead of photodetecting sensors, Tanwear \textit{et al.} used spintronic sensors to detect eye movement. Their sensor concept was based on the tunnel magnetoresistance (TMR) effect \cite{tanwear2020spintronic}. Their design integrated magnetic materials on contact lenses to detect the variation of magnetic field via spintronic sensors on the glasses frame.

\subsubsection{ERG Monitoring}
Finally, Yin \textit{et al.} demonstrated a soft graphene electrode array that was fabricated on a contact lens for measuring ERG continuously from cynomolgus monkeys \cite{yin2018soft}. Their contact lens could be used in ophthalmic diagnosis to assess if the retina works properly. The graphene sensor array ensures contact lens transparency and comfort. 

As evidenced from the literature, advances in materials design and fabrication technology lead to further innovations in sensor technology. When integrated on a soft contact lens platform, these innovative sensors widen the range of HMI applications that smart contact lenses can be used in.

\subsection{Energy}
The `Energy' module in a smart contact lens includes energy harvesting and power storage, which are two essential components for an electronic contact lens to function. To ensure uninterrupted and long term operation, contact lenses require self-sustainable and autonomous power. It can be realized by harvesting energy from the external environment. Examples of energy sources include light, heat, RF waves and motion. Additionally, after the energy being harvested, a power storage system is required to store them and provide a stable power input. Thus, the energy module of smart contact lens involves energy harvester and power storage system.

\subsubsection{Energy Harvesting}
There has been mainly three ways to harvest energy to power contact lenses: RF energy, solar energy and biochemical energy. A comparison between the amount of energy that can be harvested from these sources for wearable applications has previously been mentioned in the literature \cite{Liu2021}, with highest energy densities being achieved with photovoltaic (PV) cells. According to previous studies, there are multiple alternative energy source in smart contact lens design. Moreover, there is interest in using machine learning techniques to predict the amount of harvestable energy from these sources \cite{GhannamEnergy2021, Wahba2020}. Table III contains publications about different energy sources and the maximum power. This section involves different kinds of energy sources and storage methods being applied on contact lenses.

\begin{table}[h!]
	\small\centering
	\caption{Range of energy harvesting techniques for smart contact lenses. Highest energy densities obtained using solar (photovoltaic) energy}
    \begin{tabularx}{0.48\textwidth}{ p{2cm}|p{2.5cm}|L{3cm} }
       	\specialrule{2.5pt}{3pt}{3pt}
		Sources      & Maximum Power & References \\
		\specialrule{2.5pt}{3pt}{3pt}
		RF Energy        &  110$\mu W$   & \cite{pandey2009toward, pandey2010fully,lingley2011single,liao20113,cheng2013rectenna,hsu2015rfid,kim2015eyeglasses,chiou2017wirelessly,li2017mission,chen2018cellular,hayashi2019design,hayashi2018385mum,jeon2019143nw,chiou2019methodology,park2019printing,nasreldin2019flexible,jeon2020smart,takamatsu2020multifunctional, Xia2020}    \\ \hline Biochemical Energy      &    2.4$\mu W/cm^{2}$  & \cite{falk2012biofuel,reid2015contact,gonzalez2016transparent,frei2018power,takamatsu2020multifunctional}   \\ \hline 
		Solar Energy     &     1.24$mW/cm^{2} $    &    \cite{lingley2012contact,Xia2020,li2017mission}   \\ 
		\specialrule{2.5pt}{3pt}{3pt}
	\end{tabularx}
	\label{table:Energy Harvesting}
\end{table}

Harvesting RF energy from an external RF source is not intermittent. Thus, RF energy is a preferable and reliable energy source for smart contact lens system, as is demonstrated in fig. \ref{fig:SCL_ModuleExamplesfromLiterature}c iii. For example, a loop antenna was proposed by Takamatsu \textit{et al.} to harvest RF energy \cite{takamatsu2019highly}. Here, a external RF transmitter was placed at glass frame, which is 5 mm away from the contact lens, provided 1V DC source with up to 50\% efficiency.

Apart from RF power, researchers also investigated the possibility of integrating solar cells on contact lenses, as is shown in fig. \ref{fig:SCL_ModuleExamplesfromLiterature}c i. For example, Lingley \textit{et al.} demonstrated a contact lens with $500\times 500\times 10\mu m^{3}$ single crystal silicon solar cells \cite{lingley2012contact}. The implantable solar cell had a maximum efficiency of 1.24\% for a 725 nm source at 310 mV. Moreover, using a solar simulator that emulates an AM1.5 light source, the power density of their solar cell 1.24 $mW/cm^{2}$.  Despite this relatively high energy density for a solar cell, the actual harvestable power is small (3.1 $/mu W$) due to the constrained size of contact lenses.

Harvesting energy from the body's biochemical energy has also been demonstrated. For example, Frei \textit{et. al} demonstrated biochemical energy harvesting from metabolites in tear fluid, as shown in fig. \ref{fig:SCL_ModuleExamplesfromLiterature}c ii \cite{frei2018power}. In their design, the glucose fuel cell could only generate around 2.4$\mu W/cm^{2}$ at a cell voltage of 0.56 V. In comparison to RF and solar energy harvesting, biochemicals in tear fluid yield much lower power densities, as shown in Table \ref{table:Energy Harvesting}. 

In future, triboelectric transducers can also be considered as a feasible solution for powering electronic or smart contact lenses. These devices have not been demonstrated in the literature, but they have been used in implantable devices. With an implanted triboelectric receiver, an external acoustic wave could transfer energy remotely \cite{hinchet2019transcutaneous}. Furthermore, to overcome many of the drawbacks of each of these energy harvesters, a hybrid energy harvester can be designed for smart contact lenses \cite{Xia2020}. Such a harvester would be designed to leverage the advantages of each energy harvester. Future energy harvesting techniques may also include scavenging the eye's movements as well as blinking and winking. Here, deformation and vibration of piezoelectric materials enables kinetic energy from these gestures and movements to be converted to electrical energy \cite{markus2018piezoelectric}.

In comparison to other energy harvesting techniques, RF energy harvesting has been predominantly investigated in the literature. Although solar cells yield highest energy densities, RF harvesters are capable of scavenging more energy on the contact lens platform. Moreover, solar (or light) energy is location and time dependent, which makes it an intermittent energy source. Therefore, thanks to the advent of 5G technology and the expected proliferation of transmitter antennae, RF energy harvesters have been widely considered as a promising energy harvesting solution in smart contact lens design.

\subsubsection{Power Storage}
In the literature, there were two methods for power storage in smart contact lenses: thin film batteries and supercapacitors. 

\begin{enumerate}[i]
    \item First, there were 2 papers that demonstrated a thin film battery for power storage on contact lenses. An RF powered (flexible) thin film battery was developed by Nasreldin \textit{et al.} \cite{nasreldin2019flexible}. The electrodes of battery were were consisted of lithium nickel manganese oxide (LNMO) and Lithium Titanate (LTO), which is ring shaped with an area of 0.75 $cm^{2}$ and thickness of 100$\mu m$. It had a storage capacity of 32.25$\mu Ah$ at room temperature and can output 1.2 V to the ASIC.
 
    Lee \textit{et al.} also designed and fabricated a contact lens with a scalable thin film battery \cite{lee2018scalable}. Their battery used olivine LiFePO$_{4}$ thin film cathode and achieved a storage capacity of 35 $\mu$Wh at 0.49 $cm^{2}$ in wet conditions, which could power a glucose contact lens for 11.7 hours. Their design is demostrated in fig. \ref{fig:SCL_ModuleExamplesfromLiterature}c iv.

    \item In addition to thin film batteries, researchers considered using supercapacitors to store energy \cite{chiou2019methodology,park2019printing}. For example, Park \textit{et al.} successfully printed a supercapacitor on contact lens, which can be fully charged to 1.8 V around 240 seconds \cite{park2019printing}. The discharge time varied from 110 to 560 seconds and depended on current density. This printed supercapacitor system commonly works with a stable external RF source, which could stably operate for over 100 hours. The supercapacitor structure is shown in fig. \ref{fig:SCL_ModuleExamplesfromLiterature}c v. With rechargeable and solid-state supercapacitors, contact lenses can operate continuously with unpredictable external power sources such as RF power. In that case, such supercapacitors can provide the energy that a contact lens needs for a few minutes. 
\end{enumerate}

 Nevertheless, supercapacitors have relatively short charging and discharging times, which means that complimentary energy scavengers - such an RF converter - are necessary to ensure stable energy supply for the contact lens system. In comparison to supercapacitors, thin film batteries have higher capacities, but lower energy densities, which means they occupy a larger space. To overcome this, a hybrid solution comprising supercapacitors and thin film batteries could be developed to provide stable power supply and extended battery life. Future work on energy modules for smart contact lenses should focus on both energy harvesting and power storage to achieve self-sustainable and autonomous contact lenses.

\subsection{Driver Chip}
The driver chip is an essential part of a smart contact lens, since it is responsible for effectively managing and regulating received energy, thereby ensuring all electronic modules operate safely. This module often includes rectifiers, regulators and converters, which power the system with appropriate and reliable power signals. Moreover, signals are processed in this module via oscillators, sensor readout circuitry and logic circuitry. This is often required for reading data from the sensing module or for sending control signals to the I/O Interface. According to our review, 14.4\% of the literature's research articles involved driver chip. In fact, researchers have mainly focused on designing application-specific integrated circuits (ASICs) for health monitoring applications, as shown from our results in Table \ref{table:DriverChip}. The table lists five representative publications about ASIC design, which covers three different applications. Although there are other applications of smart contact lenses, researchers have not made any fully functional prototypes \cite{quintero2020artificial}.

\begin{table*}[h!]
	\small\centering
	\caption{Characteristics of different IC driver chip designs for various applications}
    \begin{tabularx}{\textwidth}{ p{2cm}|p{2cm}|p{2cm}|p{2cm}|p{2cm}|p{2cm}|p{2cm} }
       	\specialrule{2.5pt}{3pt}{3pt}
		Diagnostic & CMOS Process (nm)   & Frequency (MHz) & Power Consumption    & Chip Area ($mm^{2}$)  & Publication Year & Reference          \\
		\specialrule{2.5pt}{3pt}{3pt}
        IOP & 130 & 2500 & 1.4mW & 0.49 & 2010 & \cite{chow2010miniature}\\ 
            & 350 & 13.56 & 1.2mW & 2 & 2015 & \cite{donida2015circadian}\\ \hline
        Glucose & 0.13 & 1800 & 3$\mu$W & 0.22 & 2012 & \cite{liao20113}\\
            & 180 & 433 & \textless{}490nW & 2.25 & 2020 & \cite{jeon2020smart} \\
            &  65 & 4100  & 0.27nW  & 0.15  & 2018  &   \cite{hayashi2018385mum}    \\  \hline
        Tear Content & 0.18 & 920 & 110$\mu$W & 1.58 & 2017 & \cite{chiou2017towards} \\
        
		\specialrule{2.5pt}{3pt}{3pt}
	\end{tabularx}
	\label{table:DriverChip}
\end{table*}

The power consumption of smart contact lenses typically varies between 0.27 nW and 1.4 mW, depending on the function and structure of the controller chip, as shown in Table \ref{table:DriverChip} \cite{chow2010miniature,liao20113,donida2015circadian,chiou2017towards,jeon2020smart}. This energy consumption is for the whole system rather than only the driver chip. As is shown in fig. \ref{fig:SCL_ModuleExamplesfromLiterature}d, the most recent integrated circuit (IC) designed by Jeon \textit{et al.} demonstrated the possibility of developing a near zero power consumption driver, which was based on the load-shift keying (LSK) backscattering method \cite{jeon2020smart}. Consequently, this is an important step in realising autonomously powered smart contact lenses with an energy efficient driver chip. Therefore, contact lenses for glucose sensing and monitoring could be designed with less power consumption. 

The operating voltage of smart contact lenses is also a crucial parameter, since low voltages reduce power consumption and protect the eyes from potential hazards. For wearable devices, typical input voltages are 3.3 V and 5 V. However, contact lenses should operate at lower input voltage in comparison to other wearable devices since they come into direct contact with human eyes, and a minor malfunction like short circuit could cause irreversible damage. From our review, the operating voltage of contact lenses ranged from 0.165 V to 2 V. For example, Hayashi \textit{et al.} successfully designed a self-powered contact lens for glucose monitor with 0.165 V operating voltage. Their design also consumed 0.27 nW of power, which is lowest among all contact lens designs.

It is noteworthy to mention that all these articles have been published after 2010, which means that technological advances have enabled researchers to design and fabricate smart contact lenses in the lab. This highlights an emerging prospect of smart contact lens-based products in the future and the prospect of greater functionality and more powerful interface for HMI.

\subsection{Communications}
The communications block in smart contact lenses is concerned with transmitting and receiving data. For transmitter, after processing the sensor data with coder and mixer, the output signal will be transmitted by antenna. Consequently, the majority of developments in the communications module is concerned with antenna design. The important antenna design parameters for a contact lens  includes return loss, bandwidth, operating frequency, flexibility and vision obstruction.

Smart contact lenses work as an interface in HMI applications, which requires data transmission between the contact lens platform and an external processor. Thus, antennae need to be integrated on the contact lens platform \cite{yuan2020electronic}. To maximize the size of the antenna and prevent vision obstruction, researchers have often used a ring-type antenna geometry, as shown in figure \ref{fig:SCL_ModuleExamplesfromLiterature}b. 

In comparison to other antennae, the loop antenna is easier to fabricate and avoids vision obstruction, which is why it has been used in many contact lens designs \cite{takamatsu2019multifunctional,lingley2011single,chiou2015addressable}. For example, Chiou \textit{et al.} proposed a dual loop antenna design for both the receiver (Rx) and transmitter (Tx), as demonstrated in fig.\ref{fig:SCL_ModuleExamplesfromLiterature}b ii. The return loss of their design was -21 dB at 925 MHz and their -10 dB bandwidth was 21 MHz \cite{chiou2016toward}. 

From a practical and health perspective, placing an RF source 10 mm away from the contact lens seems unrealistic and hazardous, since exposure to concentrated electromagnetic radiation may cause harm to the eyes. Ting \textit{et al.} proposed loop antenna integrated on contact lens and using pig eyes to simulate practical situation and measured antenna radiation patterns, antenna gains, and radiation efficiency \cite{ting2014broadband}. Their design can cover the frequency from 1.54 to 6 GHz for ISM band applications, which is suitable for application of wireless ocular physiological monitoring.

In comparison with the loop antennae, spiral antennae demonstrate better performance, since the loop antenna is confined by horizontal polarization. Spiral antennae are more suitable as transmitter due to circular polarization, so that devices in various positions can receive the same amplitude of electric signal. Chen \textit{et al.} designed an antenna with sixteen spiral windings connected in series in a loop for near field communication, as is shown in fig.\ref{fig:SCL_ModuleExamplesfromLiterature}b iii, which can transmit data from contact lens to mobile phone \cite{chen2017warpage}. Their design has a distributed multi-layer coil structure, which could eliminate substrate warpage. Kim \textit{et al.} designed a spiral antenna with resonant frequency of 4.1 GHz, which is demonstrated in fig.\ref{fig:SCL_ModuleExamplesfromLiterature}b iv.  Communication was achieved by coupling the spiral antenna with an external antenna. With different glucose concentrations, the reflection value is different from 3.5 to 4.1 GHz, which could be received and recognized by external receiver \cite{kim2017wearable}.

\subsection{Substrate and Interconnect Technology}
Finally, all the aforementioned electronic components need to be fabricated on a particular substrate and interconnected with wires. Flexible and stretchable substrates and interconnects are therefore required to ensure user comfort. Below is a discussion regarding the materials as well as the fabrication methods and technologies used for both the contact lens substrate and its interconnects. 

\subsubsection{Substrate}
Traditionally, contact lenses were fabricated on polymer and silicone hydrogel based substrate \cite{musgrave2019contact}. On the other hand, Park \textit{et al.} proposed a novel way to fabricate stretchable contact lenses with a hybrid substrate. A photocurable optical polymer (SPC-414; EOP= 360 MPa) was photolithographically patterned with a Cu sacrificial layer, as shown in figure \ref{fig:SCL_ModuleExamplesfromLiterature}e.iii. These reinforced islands were connected by stretchable and transparent conductors \cite{park2018soft}. Moreover, Choi and Park published a concept paper regarding the use of graphene as an effective substrate material for multi-focal contact lenses \cite{choi2017smart}. They mentioned that a graphene-based contact lens could protect eyes from electromagnetic waves and dehydration. Furthermore, graphene is an extraordinarily flexibile and transparent material, which are crucial attributes of future contact lenses. 

In comparison with traditional (polymer and silicone hydrogel) contact lens substrates, future devices need better compatibility with electronic components and further investigation into encapsulation materials is necessary to prevent wearer harm from potential electronic hazards.

\subsubsection{Interconnects}
Ideally, smart contact lens interconnects need to be highly transparent, conductive, flexible, stretchable and demonstrate low sheet resistance.

Takamatsu \textit{et al.} proposed an electrochemically printed circuit on a contact lens. It was based on the polymerization of 3,4-ethylenedioxythiophene (EDOT) glue at the interface between the circuit and the hydrogel-based substrate. Their design enabled soft contact lenses to be fabricated as an interface with more complex circuits and functionality \cite{takamatsu2019highly}. Kim \textit{et al.} proposed 3D printed interconnects for smart contact lenses. They used Ag/AgNW composite ink with spray printed method to print the circuit on Polyimide substrate. Their circuit design had a small sheet resistance variation (0.396 $\Omega \cdot \square^{-1} $ ) when the device was significantly deformed. Moreover, AgNW printed circuit have relatively high conductivity and transparency, which has less heat generation and allows better vision. Their design ensured that deformation would not influence the stability of contact lens operation \cite{kim2018three}. Moreover, Alam \textit{et al.} reviewed substrate materials in previous smart contact lenses design and fabrication. They suggested that 3D printing would widely be used in next generation contact lenses. 

\subsection{Input/Output Interface}
Smart contact lenses need a mechanism to interface with the outside world. As input, we previously discussed how sensors collect data from the lens' surroundings. However, smart contact lenses may need to take action, provide feedback, or to communicate and display information back to the user. Therefore, the I/O Interface is responsible for this important functionality in smart contact lenses for HMI applications.

For example, Quintero \textit{et al.} proposed a design of an artificial iris with smart contact lens and liquid crystal material \cite{quintero2020artificial} for vision correction, as demonstrated in fig.6e i. Similar design was also made by Raducanu \textit{et al.} \cite{raducanu2020artificial} and Vanhaverbeke \textit{et al.} \cite{vanhaverbeke2017microfabrication}. They manipulate liquid crystal material by voltage variation. The switching of liquid crystal would cause the transparency variation of the contact lens, therefore it can avoid patients with iris diseases exposed to intense light. In addition to artificial iris, Milton \textit{et al.} proposed a novel contact lens design for presbyopia treatment using contact lens and liquid crystal \cite{milton2014electronic}, which is shown in fig.6e iv. Their design used the optic and electric properties of liquid crystals. The focus length of the contact lens could be controlled by voltage. Their design could help people with presbyopia could adjust the focus length of the contact lens based on their requirement.

Apart from vision correction module, there are also medical modules such as dry eye treatment module and drug delivery module. Kusam \textit{et al.} designed a dry eye treatment module on contact lens, where electroosmotic flow was generated in fluid conduit by charged hydrogel \cite{kusama2020self}. This design could help the dry eye disease treatment.

Contact lenses with safety and security module could changes color or transparency based on external environmental variation, which could be realized by liquid crystal or electrochromic material. In fig. 6e v, Kim \textit{et al.} proposed a alarm system using electrochromic material Prussian blue\cite{kim2020electrochromic}. This contact lens could turn blue automatically to warn its wearer when it detects hazards.

There are also some modules for display use. Chen \textit{et al.} proposed a concept contact lens using light-emitting diodes to achieve augmented reality, as is shown in fig. \ref{fig:SCL_ModuleExamplesfromLiterature}e vi.  Apart from academic field, a company called Mojo Vision claims to have developed a smart contact lens with display functionality using microLEDs \cite{mojovision}. They claim that their contact lens could display the size of a grain of sand to share critical information.

These modules could directly or indirectly effect on contact lens user, which is a important role in HMI application. The development of electronic devices and fabrication technology enhanced interaction between smart contact lens and user.

\subsection{Hybrids}
As previously mentioned, we categorised the literature on smart contact lenses according to six constituent building blocks. However, some references targeted more than one building block. We grouped these papers in a new block called the "hybrids". 

For example, Song \textit{et al.} designed a contact lens for both Glaucoma diagnosis and in situ drug delivery \cite{song2019multifunctional}. The contact lens stores drugs using nanopores, which could sustain the drug for up to 30 days. Therefore, this contact lens design targets both glucose sensing and drug delivery.

Chiou \textit{et al.} devised an autonomous contact lens system using RF power \cite{chiou2017wirelessly}. In their design, a radio-frequency identification (RFID) driver chip was used for capacitive sensor control and data transmission. The whole system consumed 110 $\mu W$ and was powered by external RF sources. In their design, energy can be harvested from a 30 dB RF source, which was placed 10 to 40 mm away from their contact lens \cite{chiou2016toward}. They designed the contact lens in system level, which involves communication, RFID design, energy harvesting.

Lingley \textit{et.al} devised a smart contact lens with single pixel LCD. This single pixel LCD that was automatically driven by an external RF energy source. Their design focused on both energy harvesting and information display \cite{lingley2011single}.

Kim \textit{et al.} designed a graphene-based silver nanowire (AgNW) hybrid structured sensor for glucose monitoring. The sensor used the same electrochemical principle to transform glucose concentration in tear fluid to electric signal and transmit to receiver via spiral antenna \cite{kim2017wearable}. Silver nanowires were used to improve the electrical and mechanical performance without adversely affecting the sensor's transparency and stretchability. This design involves sensing, communication and interconnection.

Raducanu \textit{et al.} designed an artificial iris system, which can change the transparency with the variation of light intensity of the environment \cite{raducanu2020artificial}. It is a systematic design which involves communication, power module, sensing, I/O interface and driver electronics.

Furthermore, Quintero \textit{et al.} devised and fabricated a hydrogel-based smart contact lens which could communicate and harvest energy from near-field source. They designed a comperhensive smart contact lens system, including communication, interconnection, energy harvesting and substrate design.


\section{Applications}
\begin{figure}[ht]
\centering
\includegraphics[scale=0.3]{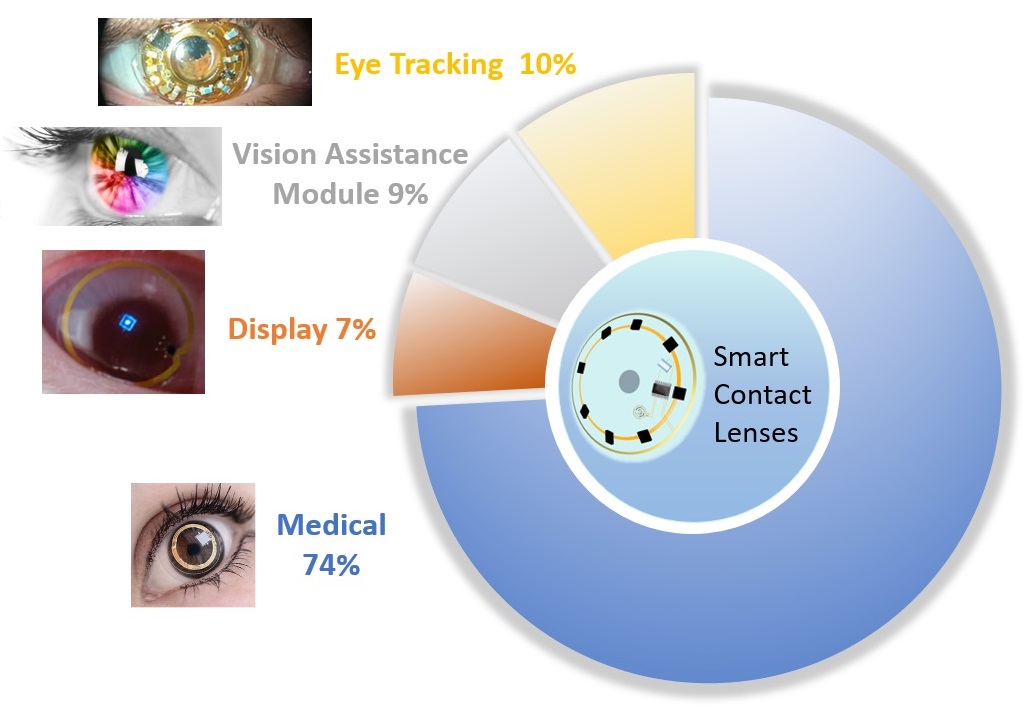}
\caption{Pie chart for smart contact lens applications. The pie chart above shows the distribution of the application areas in the reviewed manuscripts.}
\label{fig:SCL_ApplicationAreas}
\end{figure}

From our review, 58 out of our 83 manuscripts mentioned a specific application for their contact lens. The remaining 25 manuscripts were therefore discarded from the discussions below. From those 58 manuscripts, we identified four main applications areas of existing smart contact lenses, which are (i) medical, (ii) display, (iii) eye tracking and (iv) vision assistance. Moreover, the percentage of manuscripts devoted to each application area is demonstrated in fig.\ref{fig:SCL_ApplicationAreas}. 

Accordingly, the medical field accounts almost three quarters of all publications. Medical applications of smart contact lenses include the measurement of corneal temperature \cite{kinn1973liquid}, intraocular pressure (IOP) \cite{xu2016application,kingman2004glaucoma} and glucose levels \cite{kim2020wireless,lin2018noninvasive,kim2020tear}. Continuously monitoring these health indicators help improve disease diagnosis and treatment. In addition to these health indicator measurements, contact lenses have also been used for treating dry eye disease \cite{tinku2013state,kusama2020self} and for releasing drugs \cite{song2019multifunctional}.

Since 2016, smart contact lenses have been used in other non-medical applications. For example, they have been used in applications that include vision assistance, display and eye tracking. In terms of vision assistance, smart contact lenses were used for improving people's vision with ocular defects, such as iris defects \cite{quintero2020artificial} and presbyopia \cite{milton2014electronic}. Instead of static or traditional methods of vision correction, these smart contact lenses change their state based on their surrounding environmental conditions. For example, a smart contact lens alarm was demonstrated by Kim \textit{et al} \cite{kim2020electrochromic} as well as an artificial iris by Quintero \textit{et. al} \cite{quintero2020artificial} and Raducanu \textit{et al} \cite{raducanu2020artificial}.

In terms of human-machine interaction (HMI), eye tracking using smart contact lenses has also been demonstrated in the literature \cite{massin2020development,massin2020smart,tanwear2020spintronic}. Contact lens with eye tracking functionality could be used in many scenarios. For instance, in virtual gaming and online education, where there are projected benefits of contact lenses for entertainment applications in comparison to wearable devices that have limitations with accuracy and can be remedied using contact lenses \cite{massin2020smart}. Additionally, contact lens with display module are also developed to facilitate the HMI. These contact lenses could be used to transmit visualized data, which can be applied in augmented reality applications \cite{chen2020foveated}.

Moreover, contact lenses have also been demonstrated for eye tracking applications. Eye tracking is currently a hot topic that enables a computer to detect where the eye is looking. However, current methods rely on cameras and computer vision software as well as signal processing and machine learning algorithms to determine the eye's position. However, these conventional hardware configurations accommodate a limited range of head movement and suffer from poor precision \cite{brunye2019review}. In contrast, eye-wear based systems have improved accuracy, tracking speed, mobility and portability, which are important attributes for a wearable HMI application.

Furthermore, smart contact lenses were used as near eye display (NED) modules. For example, liquid crystal materials were used for displaying information, as demonstrated by the early work of de Smet \textit{et al.} in \cite{de2013progress}, as well as their most recent work in \cite{quintero2020artificial}. The versatility of such materials in contact lenses enabled them to be used for both sensing and display applications \cite{ghannam2021reconfigurable}. Moreover, the concept of using Light Emitting Diodes (LEDs) in contact lenses for augmented reality (AR) applications were proposed by \cite{chen2020foveated}. Although an experimental setup was not shown, simulation results using CODE V software predicted a high resolution smart contact lens with a wide field of view (FOV) that can move or rotate freely without the need for eye tracking. 

In terms of eye and gaze tracking applications, camera-less tracking using contact lenses was demonstrated by Massin \textit{et al.}, which was 2.5 times better than current camera-based eye-trackers \cite{massin2020development, massin2020smart}. In addition to their improved accuracy, such camera-less systems are more cost effective for eye and gaze tracking applications  \cite{liang2020live}.

\section{Challenges and Future Work}
Despite the aforementioned benefits of smart contact lenses for HMI applications, there are still many looming challenges. For example, the driving voltages required to power the various electronic modules are still large ($>1$ V). Therefore, since power consumption is high, such smart electronic platforms may be unsafe for human use on the eye. However, thanks to advancements in transistor nanofabrication, which led to a steady reduction in power supply voltage from 1.2 V in 2003 to 0.5 V in 2018, this driving voltage is expected to decrease with the advent of the 2 nm node fabrication process in 2024 \cite{weste2015cmos, ITRS2021}.

\begin{figure}[ht]
\centering
\includegraphics[scale=0.42]{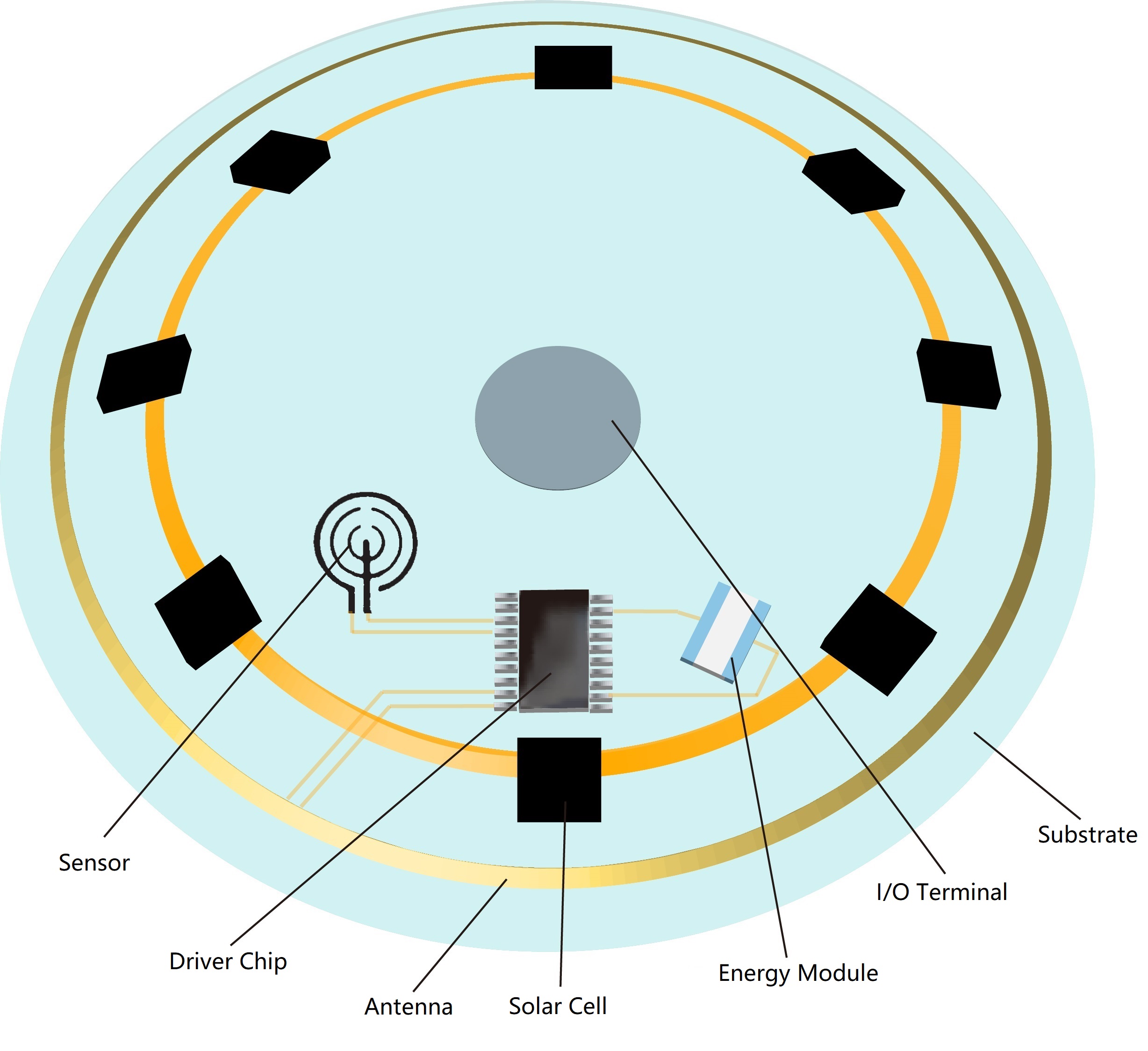}
\caption{Futuristic concept of a smart, self-powered contact lens with integrated energy harvesters, antennae, sensors, display module and power management circuitry. Such a contact lens would be a major breakthrough in Human-Machine Interaction.}
\label{fig:contactlensconcept}
\end{figure}

Moreover, since contact lenses are regarded as temporary prostheses, regulatory approval is essential before they can be used. For example, the U.S.'s Food and Drug Agency (FDA) oversees the safety, effectiveness and manufacturing of contact lenses. Such a process has been criticised as expensive and costly \cite{Van_Norman_2016}, which may hinder the introduction of smart contact lenses in large scale, especially if examination, trialling and practical training by an ophthalmologist is necessary.

Numerous studies have also shown that contact lenses may lead to serious ocular health problems if certain cleaning guidelines are poorly observed \cite{sengor2015contact, csengor2018survey}. Nevertheless, despite these challenges, surveys have shown that contact lens wearers are often more satisfied than regular eye glass wearers \cite{dias2013myopia}. Moreover, in comparison to regular eye glasses, contact lenses help widen the wearer's field of vision as well as reduce spheric and chromatic aberrations. 

Based on these developments, it is clear that the smart contact lens of the future will be autonomously driven, contain multiple sensors and will enable data processing and visualisation near the eye, as shown in figure \ref{fig:contactlensconcept}. Moreover, to ensure that such a lens is battery-less, energy will be scavenged from its surroundings using a variety of sources that could include solar cells, RF harvesters and thermoelectric generators \cite{Xia2020}.

\section{Conclusions}
Contact lenses have evolved from vision correction platforms to advanced tools for healthcare, entertainment as well as safety and security applications. This is now possible thanks to advancements in nanofabrication, materials engineering and microelectronics packaging, where electronic modules can be seamlessly integrated on flexible substrates. In all these applications, the contact lens platform plays an essential role in facilitating the interaction between humans and machines. Therefore, the purpose of this paper was to review the state-of-the-art in smart contact lens technologies and their use for Human Machine Interaction. We reviewed 83 articles published during the past 21 years which met our inclusion criteria on smart contact lenses. According to our review, the majority of research on smart or electronic contact lenses was devoted to medical applications, with almost half of all publications focusing on glucose and IOP monitoring. However, during the past five years there has been a surge in publications demonstrating the use of contact lenses for a wider range of applications, which include medical, eye tracking, vision assistance as well as displays. We discussed the feasibility of smart contact lenses as a more intelligent and powerful interface in HMI applications. We also provided an outlook into the future use of smart contact lenses in other HMI applications. Despite the challenges associated with driving voltages, as well as contact lens regulatory approvals, we believe that future trends in Very Large Scale Integration and nanofabrication will enable these electronic platforms to become a reality.

\ifCLASSOPTIONcaptionsoff
  \newpage
\fi

\bibliographystyle{IEEEtran}
\bibliography{Reference}

\begin{IEEEbiography}[{\includegraphics[width=1in,height=1.25in,clip,keepaspectratio]{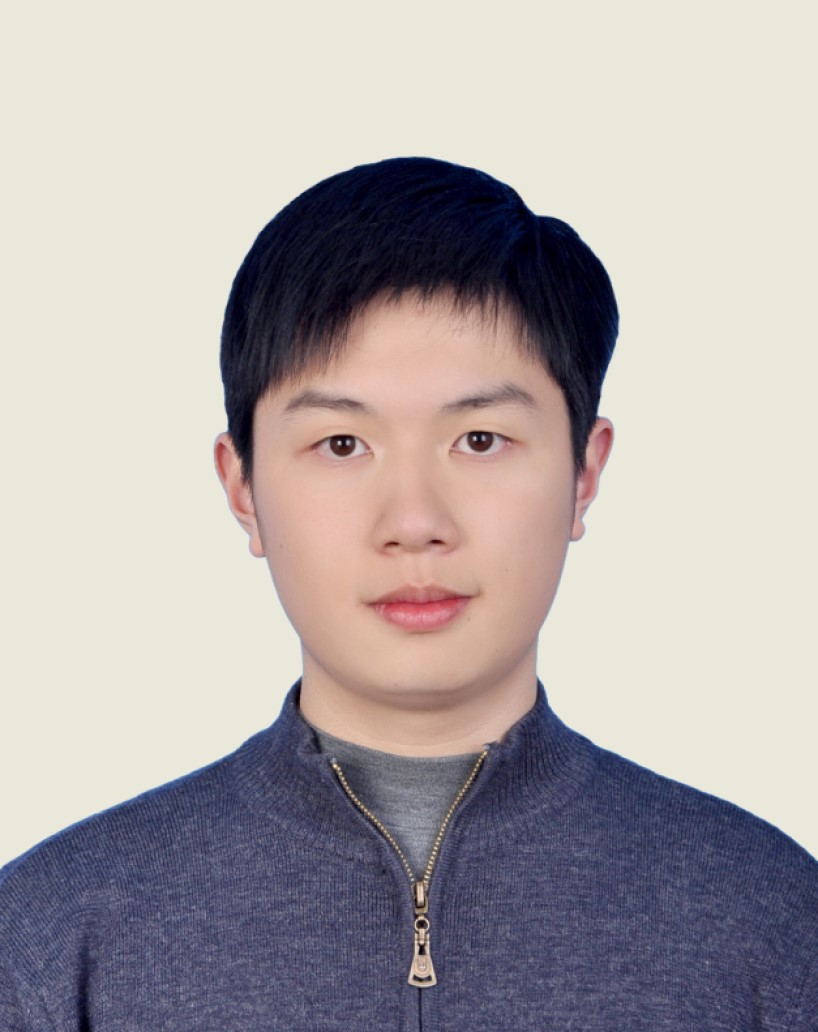}}]{Yuanjie Xia} received his BEng Degree in Electrical and Electronics Engineering at University of Electronic Science and Technology of China (UESTC) and University of Glasgow in 2020. He is currently a PhD candidate in the Microelectronics Laboratory (meLAB) at University of Glasgow and his research interests are in reconfigurable devices for HMI applications.
\end{IEEEbiography}

\begin{IEEEbiography}[{\includegraphics[width=1in,height=1.25in,clip,keepaspectratio]{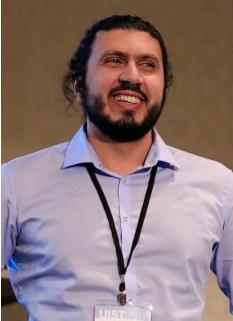}}]{Mohamed Khamis} is a Lecturer (Assistant Professor) in the School of Computing Science at the University of Glasgow, where he leads the SIRIUS Lab into research in Human-Computer Interaction, Human-centered Security and Eye Tracking. He has 90+ publications and his research received funding from several bodies including the Engineering \& Physical Sciences Research Council,  Facebook Reality Labs, the Royal Society of Edinburgh and the National Cyber Security Centre.
\end{IEEEbiography}

\begin{IEEEbiographynophoto}{Tim Wilkinson} is a Professor in Electrical Engineering at the University of Cambridge. Prof. Wilkinson has had a long term research interest into the applications of liquid crystal (LC) devices, holograms and related photonic applications.  One breakthrough made recently was in algorithms which lead to the commercialisation of holographic projectors. Core to his research has been liquid crystal device fabrication and the search for the optimal materials for different applications. 
\end{IEEEbiographynophoto}

\begin{IEEEbiographynophoto}{Haider Butt} is an Associate Professor in Mechanical Engineering at Khalifa University. He was awarded both an M.Phil and a PhD in Electrical Engineering from the University of Cambridge in 2008 and in 2012. He has published over 100 peer-reviewed articles in prestigious journals including Nature Communications, Advanced Materials, Biotechnology Advances, Light Science and Applications, and ACS Nano. He has earned international recognition for his research and leadership in optical sensors where he has contributions to nanoscale devices by conceiving novel holographic lithography methods to produce optical transducers. His most pioneering works include glucose sensing contact lenses, contact lenses for color blind patients and carbon nanotube holograms.
\end{IEEEbiographynophoto}

\begin{IEEEbiography}[{\includegraphics[width=1in,height=1.25in,clip,keepaspectratio]{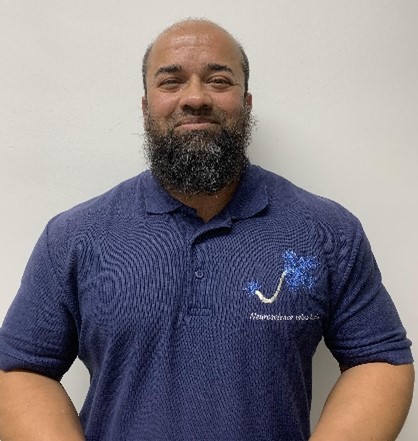}}]{Zubair Ahmed}
is a Professor in Neuroscience in the Institute of Inflammation and Ageing at the University of Birmingham. He received his PhD in 1999 from University College London and held positions at the Institute of Neurology in London before joining Birmingham in 2002. He held a prestigous RCUK Academic Fellowship, before being promoted to Lecturer and recently to Full Professor in Neuroscience. He is currently the Lead for Neuroscience and Ophthalmology and his research interests include the definition of the molecular mechanisms underpinning ocular, spinal and brain injuries. 
\end{IEEEbiography}

\begin{IEEEbiography}[{\includegraphics[width=1in,height=1.25in,clip,keepaspectratio]{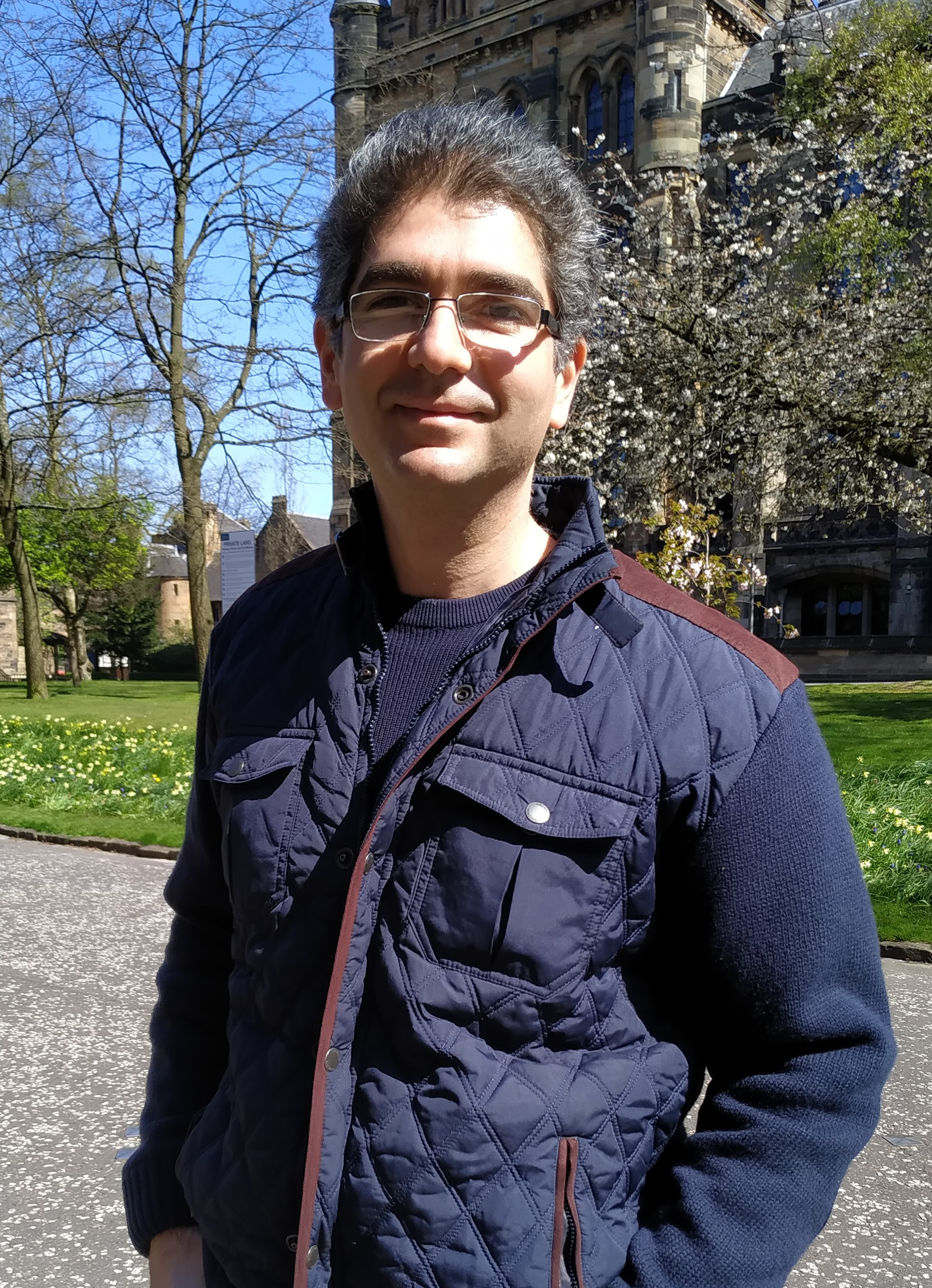}}]{Rami Ghannam} (SM'18) is a Senior Lecturer (Associate Professor) in Electronic Engineering at the University of Glasgow. He received his BEng degree in Electronic Engineering from King's College London and was awarded the Siemens Prize. He subsequently received his DIC and MSc degrees from Imperial College London, as well as his PhD from the University of Cambridge in 2007. He held previous industrial positions at Nortel Networks and IBM Research GmbH. His research interests are in Photonics, Sensors and Wearable Computing. He is a Senior Member of the IEEE, a Senior Fellow of Glasgow’s Recognising Excellence in Teaching scheme and serves as Scotland’s Regional Chair of the IEEE Education Society.
\vspace{-10 mm}
\end{IEEEbiography}

\end{document}